\documentclass[numberedappendix, appendixfloats]{emulateapj}
\usepackage{longtable, hyperref, graphicx, subfigure}
\usepackage[fleqn]{amsmath}
\usepackage{lmodern}
\usepackage[T1]{fontenc}
\usepackage{url}
\usepackage{todonotes}
\newcounter{column_number}
\newcommand{\Msun}{\ifmmode {M_{\odot}}\else{M$_{\odot}$}\fi}
\newcommand{\lapprox }{{\lower0.8ex\hbox{$\buildrel <\over\sim$}}}
\newcommand{\gapprox }{{\lower0.8ex\hbox{$\buildrel >\over\sim$}}}

\def\asec{\ifmmode^{\prime\prime}\else$^{\prime\prime}$\fi}

\def\LX{$L_{\mathrm{X}}$}
\def\Ha{$\mathrm{H}\alpha$}
\def\LHa{$L_{\mathrm{H}\alpha}$}
\def\LLX{$L_{\mathrm{X}}/L_{\mathrm{bol}}$}
\def\LLH{$L_{\mathrm{H}\alpha}/L_{\mathrm{bol}}$}
\def\Lbol{$L_{\mathrm{bol}}$}
\def\Prot{$P_{\mathrm{rot}}$}
\def\Pmem{$P_{\mathrm{mem}}$}

\slugcomment{DRAFT \today}
\shorttitle{Chromospheres and Coronae in M37}
\shortauthors{N\'u\~nez, Ag{\"u}eros, Covey, \& L\'opez-Morales}

\begin{document}

\title{Chromospheric and Coronal Activity in the 500-Myr-old Open Cluster M37: Evidence for Coronal Stripping?} 
\author{Alejandro N\'u\~nez\altaffilmark{1},
Marcel~A.~Ag{\"u}eros\altaffilmark{1}, Kevin R.~Covey\altaffilmark{2}, and Mercedes L\'opez-Morales\altaffilmark{3}
}

\altaffiltext{1}{Department of Astronomy, Columbia University, 550 West 120th Street, New York, NY 10027, USA}
\altaffiltext{2}{Department of Physics and Astronomy, Western Washington University, Bellingham WA 98225, USA}
\altaffiltext{3}{Harvard-Smithsonian Center for Astrophysics, 60 Garden Street, Cambridge, MA 02183, USA}

\begin{abstract}  
We present the results of a spectroscopic survey to characterize chromospheric activity, as measured by \Ha\ emission, in low-mass members of the 500-Myr-old open cluster M37. Combining our new measurements of \Ha\ luminosities (\LHa) with previously cataloged stellar properties, we identify saturated and unsaturated regimes in the dependence of the \LHa-to-bolometric-luminosity ratio, \LLH, on the Rossby number $R_o$. All rotators with $R_o$ smaller than 0.03$\pm0.01$ converge to an activity level of \LLH\ = $(1.27$$\pm$$0.02)\times10^{-4}$. This saturation threshold ($R_{o,\mathrm{sat}}=0.03$$\pm$0.01) is statistically smaller than that found in most studies of the rotation-activity relation. In the unsaturated regime, slower rotators have lower levels of chromospheric activity, with \LLH($R_o$) following a power-law of index $\beta=-0.51$$\pm$0.02, slightly shallower than the one found for a combined $\approx$650-Myr-old sample of Hyades and Praesepe stars. By comparing this unsaturated behavior to that previously found for coronal activity in M37 (as measured via the X-ray luminosity, \LX), we confirm that chromospheric activity decays at a much slower rate than coronal activity with increasing $R_o$. While a comparison of \LHa\ and \LX\ for M37 members with measurements of both reveals a nearly 1:1 relation, removing the mass-dependencies by comparing instead \LLH\ and \LLX\ does not provide clear evidence for such a relation. Finally, we find that $R_{o,\mathrm{sat}}$ is smaller for our chromospheric than for our coronal indicator of activity ($R_{o,\mathrm{sat}}=0.03$$\pm$0.01 versus 0.09$\pm$0.01). We interpret this as possible evidence for coronal stripping.
\end{abstract}


\section{Introduction}
In late-type, main-sequence stars, the rotation rate and the strength of the magnetic field decrease over time \citep{skumanich72,palla81}. This is thought to result from a feedback loop in which winds carry angular momentum away from the star, braking its rotation and diminishing the shear between the internal radiative and convective zones, which is responsible for generating the magnetic field \citep{Parker93}. The resulting, weaker magnetic field then produces weaker winds; these continue to spin down the star and further weaken its magnetic field, but at a diminished rate.

This relationship between age, rotation period (\Prot), and magnetic activity has been modeled empirically with data from the homogeneous, co-eval populations of open clusters. Two of the commonly used tracers of stellar activity are X-ray flux, $f_X$, which in late-type stars originates in the corona \citep{Vaiana81}, and \Ha\ emission, which originates in the chromosphere \citep{Campbell1983}. Due to their linked heating mechanisms, a correlation is expected between X-ray and H$\alpha$ emission in magnetically active stars.

A well-calibrated age-rotation-activity relation (ARAR) would be particularly valuable for low-mass stars. Earth-like planets are most likely to be discovered in the habitable zones of nearby, old, low-mass, field stars \citep[cf.~discovery of a 1.3 M$_\oplus$ planet around Proxima Centauri;][]{Anglada2016}. Understanding these planets' radiation environments and potential habitability demands a robust ARAR that could be applied to their parent stars. If we knew the dependence of \Prot, $f_X$, or H$\alpha$ on age, a measurement of one of these quantities could be used to determine an accurate age for any isolated field star. Currently, however, we invert the process, adopting canonical ages for field stars to constrain the behavior of the ARAR at the oldest ages \citep[e.g.,][]{Kiraga2007, irwin2011}.

Few open clusters have been systematically surveyed for the three quantities \Prot, $f_X$, and H$\alpha$, and moreover there is a scarcity of accessible clusters older than $\approx$150~Myr. The Hyades and Praesepe (both $\approx$650 Myr) serve as the only anchors for our understanding of the dependence of stellar activity on rotation between a few 100 Myr and the age of field stars ($\gapprox$2~Gyr). Unfortunately, the \Ha--\Prot\ and $f_X$--\Prot\ relations do not agree for stars in these two clusters: compared to the H$\alpha$-to-bolometric luminosity ratio (\LLH), the X-ray-to-bolometric luminosity ratio (\LLX) depends much more strongly on \Prot\ for slow rotators \citep[][hereafter D14]{Douglas14}. The exact relationship between rotation and activity tracers in a single-aged population remains elusive. Furthermore, the same behavior has been observed in mixed-age samples where different tracers of magnetic activity ($f_X$, \Ha, and Ca\textsc{ii}) are compared \citep[e.g.,][]{rauscher2006, stelzer2013}. To understand the physical underpinnings of the ARAR we need to be confident that we understand how different age indicators evolve, and why.

The $\approx$500-Myr-old open cluster M37 \citep[NGC~2099, 1490$\pm$120 pc;][hereafter H08]{Hartman2008}, has been extensively surveyed for \Prot\ \citep{messina08a,Hartman09} and for $f_X$ \citep[][hereafter Paper I]{Nunez2015}. With more than 400 cluster members with \Prot\ and more than 270 with $f_X$ measurements, M37 is the best laboratory for comparing the behavior of H$\alpha$ emission and of $f_X$ and their relation to \Prot\ in a single-aged population, as there is no other cluster older than the Pleiades \citep[$\approx$112 Myr,][]{stauffer1998} with comparable \Prot\ and X-ray coverage and as rich a membership \citep[cf.~analysis of six other clusters in][]{Nunez2016}.

Here we complement our study in Paper I of the relationship between \Prot\ and $f_X$ in M37 with an examination of the relationship between \Prot\ and \Ha\ emission, and therefore between these two tracers of magnetic activity at 500 Myr. To measure \Ha\ emission, we obtain optical spectra of 298 M37 members with the 6.5-m telescope at the MMT Observatory, Mt.~Hopkins, AZ; 125 of our targets show the line in emission.\footnote{Observations reported here were obtained at the MMT Observatory, a joint facility of the University of Arizona and the Smithsonian Institution.}

In Section \ref{data}, we describe how we assembled our cluster membership catalog, our spectroscopic data, and the \Prot\ data we collected from the literature. In Section \ref{properties}, we characterize stars in M37 using available photometry and our spectra. We present our results in Section \ref{results} before concluding in Section \ref{summary}. 

\section{DATA}\label{data}
\subsection{Cluster Members}
H08 obtained {\it g'r'i'} images with Megacam on the 6.5 m MMT telescope of a 24$\arcmin$$\times$24$\arcmin$ area centered on M37. These authors then converted the instrumental magnitudes to Sloan Digital Sky Survey \citep[SDSS;][]{york00} $gri$ magnitudes for $\approx$16,500 objects with measurements in all three bands.\footnote{H08 observed the equatorial Sloan field centered at $03^{\mathrm{h}}20^{\mathrm{m}}00^{\mathrm{s}}$, $00^{\circ}00'00''$ (J2000) to constrain air mass, and then parametrized the instrumental-to-SDSS magnitude relation using a list of observed stars matched to stars extracted from the SDSS Data Release 5 \citep{dr5}.} The typical one-sigma ($\sigma$) error is 0.01 mag, and the magnitude coverage is $10\ \lapprox\ r\ \lapprox\ 25$~mag. 

In Paper I  we used this {\it gri} photometry and the distance from the cluster core to identify cluster members. Each star was assigned a probability of being a single member ($P_{\mathrm{s}}$), a likely binary member ($P_{\mathrm{b}}$), or a field star ($P_{\mathrm{f}}$), with $P_{\mathrm{s}}$ + $P_{\mathrm{b}} \equiv$ \Pmem\ and $P_{\mathrm{s}}$ + $P_{\mathrm{b}}$ + $P_{\mathrm{f}}$ = 1 for each star (see section 3.3 and appendix A of Paper I for details). 

We identify 1643 stars with \Pmem\ $\geq$ 0.2, which we use as the \Pmem\ cutoff for cluster membership. This low \Pmem\ threshold may result in our catalog being significantly contaminated by field stars. However, we expect this contamination to be minimal when considering stars with \Prot\ and activity measurements, as field stars tend to be much older and, therefore, very slow rotators unlikely to be detected in the X ray or to have \Ha\ emission.

With this in mind, we also search for stars with 0.1~$\leq$~\Pmem\ <~0.2 that have H$\alpha$ emission in our spectra (see Section~\ref{spectroscopy}). We found 27 such stars. These stars' location in a color-magnitude diagram (CMD) and a mass--\Ha\ equivalent width (EqW) plot indicates that their properties are consistent with those of cluster members, and we therefore classify these stars as {\it bona fide} M37 members and assign them a \Pmem\ $=$ 999.

Member stars with $P_{\mathrm{b}}$ > $P_{\mathrm{s}}$ are flagged as likely cluster binaries. As described in Paper I, $P_{\mathrm{b}}$ is solely based on photometric distance from the main sequence in the ($i$, $g-i$) CMD.

\citet{messina08a} surveyed M37 in the optical with the 1 m telescope at the Mt.~Lemmon Optical Astronomy Observatory, AZ. We use these data to determine whether stars without a counterpart in the H08 survey should be included in our M37 catalog. We determine their membership by visually inspecting their location in the ($i$, $g-i$) CMD, using their $BV$ photometry transformed into $gi$.\footnote{We used the transformations derived by \citet{jester2005} and in \url{http://www.sdss.org/dr12/algorithms/sdssUBVRITransform}.} Of the stars identified as cluster members using this approach, we include in our analysis five that have \Prot\ and that have either an X-ray counterpart in Paper I or \Ha\ in emission in our Hectospec spectra. We also assign them \Pmem~$= 999$.

Table~\ref{tbl:columns} describes the 21 columns in our resulting catalog of cluster members, and Table~\ref{tbl:clusterstars} shows some of these columns for the first five stars in the catalog. This table is available online in its entirety. 

\begin{deluxetable}{@{}ll}
\tabletypesize{\scriptsize} 

\tablecaption{Description of Columns in M37 Cluster Catalog \label{tbl:columns}}

\tablehead{
\colhead{Col.} & \colhead{Description} \\[-0.1 in]                
\setcounter{column_number}{1}
}

\startdata
1      & Source ID from H08 (>10000) or \citet{messina08a}\\
       & (<10000). \\
2, 3   & Right ascension and declination of object (J2000).\\
4      & Membership probability. \Pmem\\
5      & Binary flag from Paper I: 0, likely single star;\\
	   & 1, likely binary.\\
6-8    & $giJ$ magnitudes. \\
9      & Stellar mass. \\
10     & Rossby number $R_o$. \\
11,12  & EqW and standard deviation of the H$\alpha$ line.\\
13-16  & EqW and standard deviation of the [N\textsc{ii}]$\lambda$6584 and \\
	   & [N\textsc{ii}]$\lambda$6548  lines.\\
17     & Empirical ratio $\chi$ of the continuum flux near the H$\alpha$ \\
	   & line and the apparent bolometric flux.\\
18     & H$\alpha$-to-bolometric luminosity ratio \LLH.\\
19     & Standard deviation of \LLH.\\
20     & X-ray-to-bolometric luminosity ratio \LLX.\\
21     & Standard deviation of \LLX.\\[-0.1 in]
\enddata

\end{deluxetable}

\begin{deluxetable*}{cccccccccrrrr}
\tabletypesize{\scriptsize} 

\tablecaption{M37 Members \label{tbl:clusterstars}}

\tablehead{
\colhead{Name} & \colhead{$\alpha$ (J2000)} & \colhead{$\delta$ (J2000)} & \colhead{\Pmem} & \colhead{Bin.} & \colhead{$g$} & \colhead{$i$} & \colhead{Mass} & \colhead{Ro} & \colhead{EqW H$\alpha$} & \colhead{$\chi$} & \colhead{\LLH} & \colhead{\LLX} \\  
\colhead{} & \multicolumn{2}{c}{(\arcdeg)} & \colhead{} & \colhead{flag} & \colhead{(mag)} & \colhead{(mag)} & \colhead{(\Msun)} & \colhead{} & \colhead{(\AA)} & \colhead{} & \colhead{} & \colhead{} \\
\colhead{(1)} & \colhead{(2)} & \colhead{(3)} & \colhead{(4)} & \colhead{(5)} & \colhead{(6)} & \colhead{(7)} & \colhead{(9)} & \colhead{(10)} & \colhead{(11)} & \colhead{(17)} & \colhead{(18)}  & \colhead{(20)} \\[-0.1 in]
}
\startdata
140028 & 88.157125 & 32.538739 & 0.77 & 0 & 17.09 & 15.97 & 1.04 & .. & 2.01 & .. & .. & 2.21E$-$05 \\
140036 & 88.158163 & 32.570939 & 0.59 & 0 & 17.48 & 16.23 & 1.01 & 0.08 & 1.25 & 9.26E$-$05 & .. & 9.18E$-$04 \\
150134 & 88.186100 & 32.491900 & 0.36 & 0 & 20.40 & 18.05 & 0.63 & 0.05 & $-$1.89 & 6.64E$-$05 & 1.26E$-$04 & 1.31E$-$03 \\
240167 & 88.031825 & 32.524308 & 0.88 & 0 & 20.94 & 18.46 & 0.61 & .. & $-$0.70 & 6.59E$-$05 & 4.62E$-$05 & 7.46E$-$04 \\
230343 & 88.023129 & 32.546933 & 0.49 & 1 & 23.05 & 19.80 & 0.39 & 0.08 & $-$1.25 & 2.75E$-$05 & 1.19E$-$05 & ..
%20040 & 88.238650 & 32.688167 & 0.38 & 0 & 18.54 & 17.04 & 0.83 & 0.16 & 1.13 & 9.62E-05 & 0.00E+00 & 0.00E+00 \\
%20101 & 88.294533 & 32.682731 & 0.26 & 0 & 21.23 & 18.60 & 0.60 & 0.39 & 0.47 & 5.22E-05 & 0.00E+00 & 0.00E+00 \\
%30145 & 88.260050 & 32.651308 & 0.34 & 0 & 21.47 & 18.73 & 0.60 & 0.33 & 0.00 & 4.75E-05 & 0.00E+00 & 0.00E+00 \\
%40038 & 88.256183 & 32.579903 & 0.37 & 0 & 18.29 & 16.82 & 0.88 & 0.40 & 1.44 & 9.41E-05 & 0.00E+00 & 0.00E+00 \\
%40039 & 88.233092 & 32.593967 & 0.53 & 0 & 18.33 & 16.85 & 0.88 & 0.38 & 1.27 & 9.29E-05 & 0.00E+00 & 0.00E+00
\enddata

\tablecomments{This table is available in its entirety in the electronic edition of the \apj. Some columns and rows are shown here for guidance regarding its form and content. Table \ref{tbl:columns} describes all the columns in this table.}

\end{deluxetable*}

\subsection{Spectroscopy}\label{spectroscopy}
We obtained spectra of M37 stars with Hectospec \citep{Fabricant2005} on the MMT 6.5 m telescope. We used five different fiber configurations, set up using the program \texttt{XFITFIBS}, centered near $\alpha=05^{\mathrm{h}}52^{\mathrm{m}}18^{\mathrm{s}}$, $\delta=+32^{\circ}33'00\farcs9$ (J2000), to observe 356 stars in the field of view of M37.

\begin{figure*}
\centerline{\includegraphics{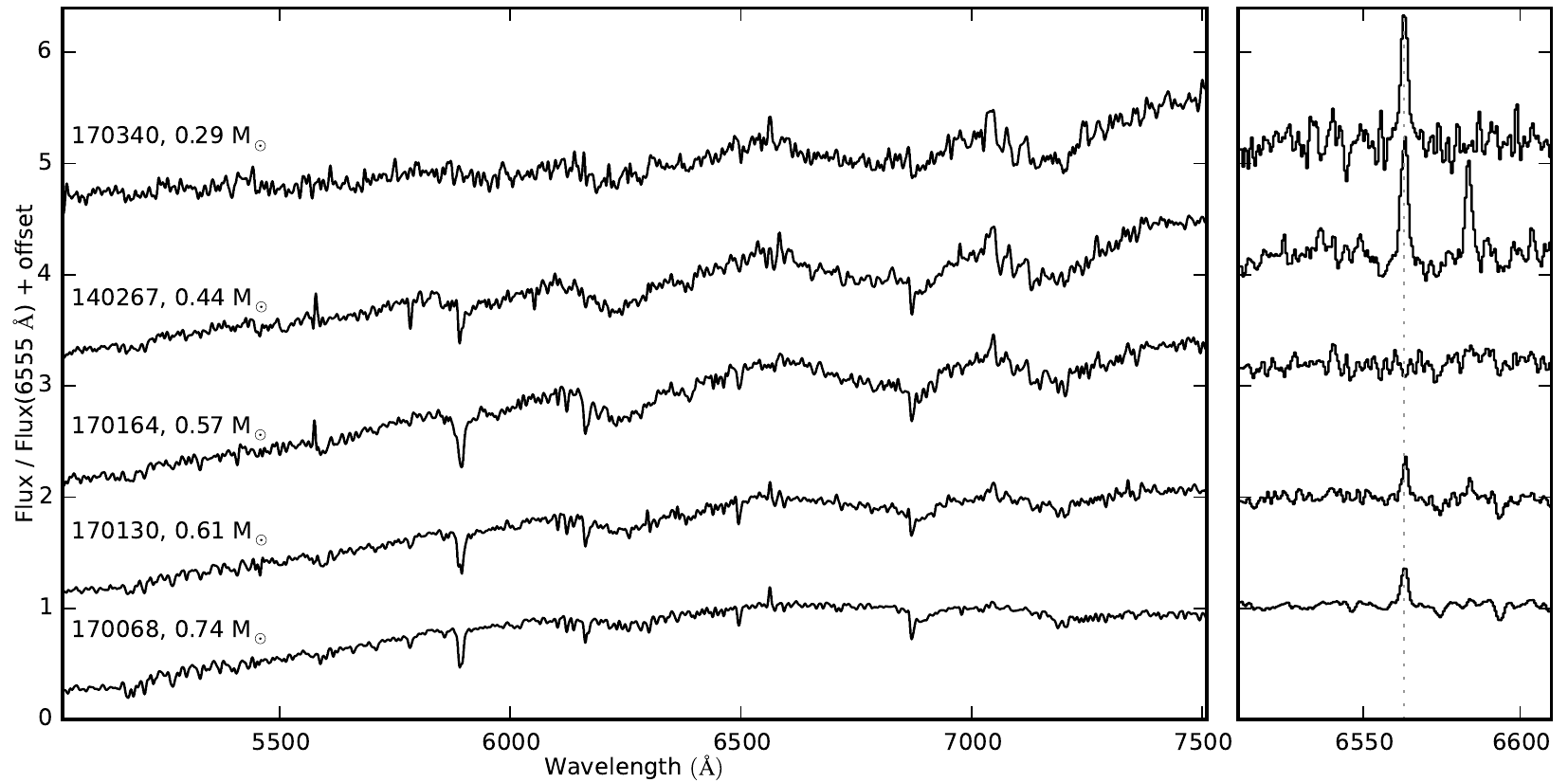}}
\caption{Five representative MMT Hectospec spectra of M37 stars, labeled with their H08 designations and photometrically derived masses. Each spectrum is normalized to the flux at 6555~\AA\ and smoothed using a 20 pixel smoothing window. The right panel shows a close-up of the \Ha\ line, with the spectra smoothed using a four pixel smoothing window. The vertical dotted line indicates the \Ha\ line.}
\label{fig:spectra}
\end{figure*}

We used the 600 line grating, which has a central wavelength of 6500~\AA\ and a free spectral range of 5770~\AA, covering the \Ha\ line with a spectral resolution $R \approx 1000$. Our targets lie in the range $15.1 < r < 22.6$~mag. Stars with $r \gapprox\ 20$~mag typically were observed for 300--350 min; stars with $17\ \lapprox\ r\ \lapprox\ 20$, for 100--300 min; and stars with $r\ \lapprox\ 17$, for $\approx$90 min. The median signal-to-noise of our spectra at the \Ha\ line core (6563~\AA) is 44. Table~\ref{tbl:observations} summarizes our observations, and Figure~\ref{fig:spectra} shows five example spectra, illustrating the varying strength of the \Ha\ line seen in our sample.

The data were reduced automatically by the Telescope Data Center using the HSRED v2.0 pipeline. HSRED performs the basic reduction tasks: bias subtraction, flat-fielding, arc calibration, and sky subtraction.\footnote{See \url{http://mmto.org/~rcool/hsred/index.html} for a  description of HSRED.}

\begin{deluxetable}{@{}rccccc@{}}
\centering 
\tabletypesize{\scriptsize}

\tablecaption{ Log of MMT Hectospec Observations
 \label{tbl:observations}}

\tablehead{
\colhead{Date} &
\colhead{C\tablenotemark{a}} &  
\colhead{Exp.} & 
\multicolumn{2}{c}{Nominal Aimpoint} & 
\colhead{Num. of}\\
\cline{4-5}
\colhead{} &
\colhead{} &  
\colhead{(s)} & 
\colhead{$\alpha_{\rm J2000}$} & 
\colhead{$\delta_{\rm J2000}$} & 
\colhead{Spectra}
}

\startdata
2015-02-18 & 2 & 3840 & 05:52:18.91 & +32:31:54.16 & 134 \\
2015-02-19 & 2 & 2810 & 05:52:18.91 & +32:31:54.16 & 132 \\
2015-02-19 & 3 & 2400 & 05:52:18.64 & +32:33:05.93 & 121 \\
2015-02-20 & 3 & 3060 & 05:52:18.64 & +32:33:05.93 & 123 \\
2015-04-18 & 1 & 5400 & 05:52:11.84 & +32:32:45.91 & 105 \\
2015-09-20 & 5 & 3600 & 05:52:17.96 & +32:31:37.13 & 80  \\
2015-11-21 & 5 & 8400 & 05:52:17.96 & +32:31:37.13 & 80  \\
2015-11-22 & 4 & 3600 & 05:52:22.40 & +32:31:49.42 & 103
\enddata
\tablenotetext{a}{Configuration number.}

% configuration 4 == 21n; configuration 5 == 31n

\end{deluxetable}

\subsection{Rotation Periods}\label{rotations}
\citet{messina08a} and \citet{Hartman09} published independent surveys of stellar rotation in M37, which we consolidated in Paper I (see sections 3.1 and 3.2 of Paper I for details). Briefly: \citet{messina08a} used both the Scargle-Press and CLEAN periodogram techniques to measure variability for stars in the range 13 < $V$ < 20 mag and reported \Prot\ values for 120 stars. 

\citet{Hartman09} applied the multiharmonic analysis of variance algorithm of \citet{schwarzenberg96} to light curves for 15 < $r$ < 23 mag stars included in the \citet{hartman08} list of variable stars. For 372 stars, these authors found that \Prot\ differed <10\% when they re-calculated it using only one, two, or three terms of the harmonic series fitted to the light curves. For these stars, they chose \Prot\ calculated using two terms and reported them as ``clean'' \Prot\ values. As we noted in Paper I, non-clean \Prot\ values add unnecessary ambiguity to our analysis. We therefore include non-clean rotators only if they also have a \Prot\ from \citet{messina08a} and the two \Prot\ measurements do not differ by more than 20\%.

We matched rotators from both surveys using a 10\asec\ matching radius. Individual inspections of the matches comparing the $gri$ photometry from H08 with the $BV$ photometry from \citet{messina08a} then determined the most likely correct match. The liberal matching and individual inspection was necessary because the astrometry of objects in \citet{messina08a} displays a widespread non-linear distortion.\footnote{See section 3.2 of Paper I for details.} For stars with \Prot\ measurements in both surveys, we adopted the \Prot\ from \citet{Hartman09}. Our consolidated list of rotators includes 657 stars with at least one \Prot\ measurement, 426 of which have \Pmem\ $\geq$ 0.2.

\section{Deriving Stellar Properties}\label{properties}

\subsection{Calculating Masses and \Lbol}\label{masses}
We estimate masses for M37 members using the mass-absolute $r$ magnitude ($M_r$) relation of \citet{Kraus2007}, who generated empirical spectral energy distributions for B8-L0 stars that are calibrated using the 650-Myr-old Praesepe cluster with SDSS $ugriz$ and Two Micron All Sky Survey \citep[2MASS,][]{2mass} $JHK$ photometry (see section 4.1.1 of Paper I). To obtain $M_r$ we estimate the total absorption in $r$ ($A_r$) using the extinction tables of \citet{schlafly2011} assuming $R_V$ = 3.1 and adopting a reddening $E$($B-V$) = 0.227 and distance 1490 pc.

Similarly, we estimate \Lbol\ for M37 members by using the effective temperature-$M_r$ relation of \citet{Kraus2007}. After obtaining an effective temperature for each star from $M_r$, we use the corresponding bolometric correction in the \citet{Girardi04} tables, which we tailor to the SDSS filter system. This allows us to calculate bolometric magnitudes and luminosities, the latter by again using the distance of 1490 pc to the cluster.

\subsection{Measuring \Ha\ EqWs and Obtaining \LLH}\label{Halpha}
To obtain \Ha\ EqW measurements for our spectra, we use the \texttt{PHEW} tool,\footnote{See \url{https://zenodo.org/record/47889\#.V5hjMlcZZtI}.} which automates the EqW calculation using \texttt{PySpecKit} \citep{Ginsburg2011} and performs Monte Carlo iterations to obtain EqW uncertainties.

We set 6563~$\AA$ as the center of the \Ha\ line and define the continuum flux for each spectrum as the average flux between 6540$-$6558 and 6572$-$6590 \AA. This is modified by eye when the line is broad or shifted away from 6563~$\AA$ (e.g., because of binarity or in high radial-velocity objects). In cases where we had more than one spectra for the same star, we weighted-mean combine the spectra before performing the EqW measurement. Lastly, we perform 1000 Monte Carlo iterations to calculate 1$\sigma$ uncertainties on the EqW measurements. In each iteration, we varied the spectral flux at each pixel by adding a fraction ---pulled randomly from a Gaussian distribution--- of the intrinsic flux uncertainty. The standard deviation of the 1000 measured EqW values corresponds to the 1$\sigma$ uncertainties that we report. We measure \Ha\ EqW for 294 cluster members; 125 (including 12 likely binaries) show the line in emission.

\begin{figure}
\includegraphics{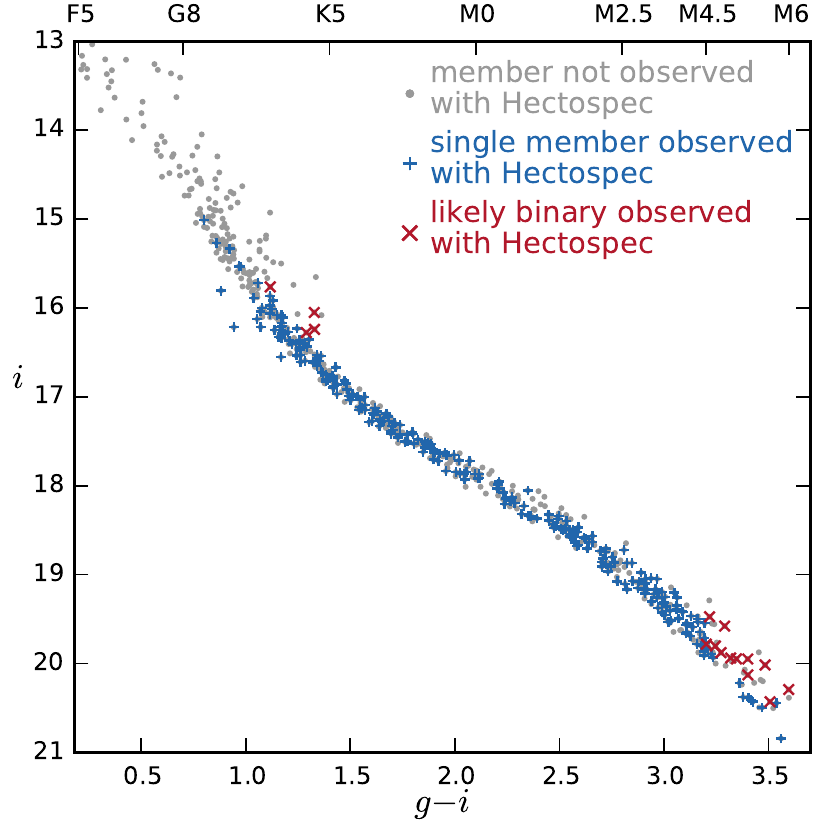}
\caption{CMD for stars in the field of view of M37 with \Pmem~$\geq 0.2$ in the color range $0.2<g-i<3.7$. Targets of our MMT Hectospec observations are highlighted in blue (single members) and red (photometrically identified likely binaries).}
\label{fig:CMD}
\end{figure}

Figure~\ref{fig:CMD} shows a color-magnitude diagram with stars in the color range $0.2< (g-i) <3.7$ having \Pmem~$\geq 0.2$. Highlighted in blue (single members) and red (likely binary members) are stars for which we obtain an EqW measurement; negative EqW values correspond to emission. For a handful of stars, our spectra are too noisy to perform a measurement.

\begin{figure}
\includegraphics{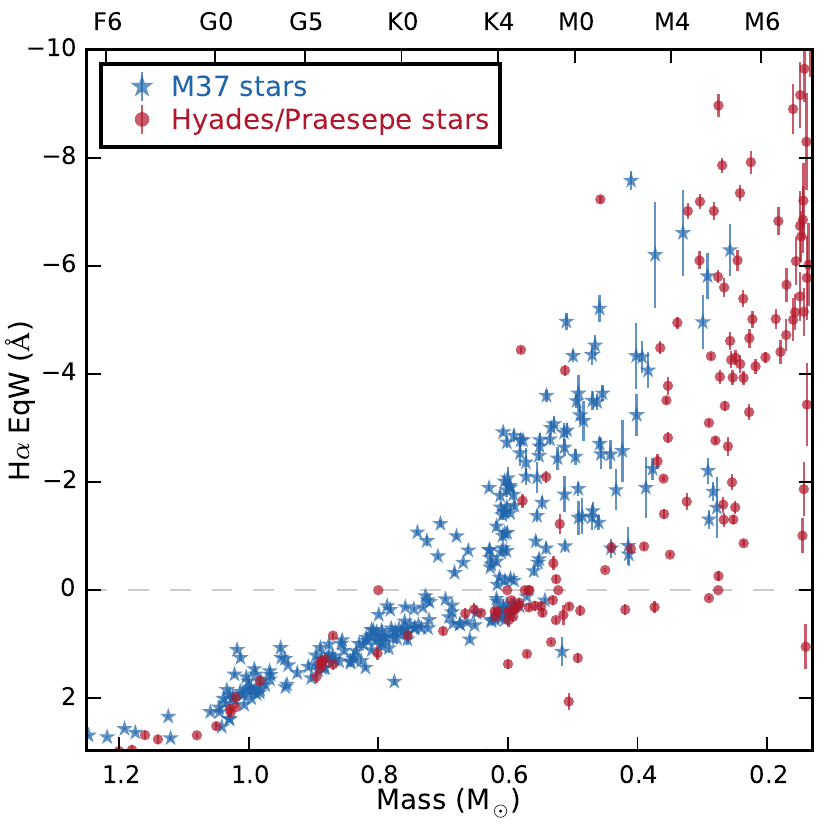}
\caption{H$\alpha$ EqW as a function of mass for M37 stars (blue stars symbols) and HyPra stars (red circles). All masses are derived using the absolute magnitude-mass relation of \citet{Kraus2007} using SDSS or SDSS-equivalent $r$ magnitudes. All EqW measurements are performed using the \texttt{PHEW} tool. M37 stars identified as likely binaries by \citet{Nunez2015} and HyPra stars identified as candidate or confirmed binaries by D14 and \citet{Douglas2016} are excluded from the plot. \Ha\ activity is evident in higher-mass stars in M37 than in HyPra, indicating that M37 is indeed the younger cluster.}
\label{fig:mass_EqW}
\end{figure}

D14 studied \Ha\ activity for stars in the Hyades and Praesepe. These two benchmark open clusters have similar ages ($\approx$650 Myr) and metallicities, and D14 merged their membership catalogs to create the so-called HyPra cluster. We compare \Ha\ activity in M37 and in HyPra in Figure~\ref{fig:mass_EqW}. We obtain SDSS $r$ magnitudes for HyPra stars (using a 1$\arcsec$ matching radius) to calculate stellar masses using the same \citet{Kraus2007} mass-$M_r$ relation we use for M37 stars. Only 20\% of HyPra stars in Figure~\ref{fig:mass_EqW} do not have SDSS counterparts; for those we use the stellar masses from D14, which were derived using 2MASS $K$ magnitudes and the \citet{Kraus2007} mass-absolute $K$ magnitude relation.\footnote{Our $M_r$-derived masses for HyPra stars are on average 7\% smaller than the $M_K$-derived masses given in D14.} We also perform our own EqW measurements for the spectra of these HyPra stars. We exclude from this comparison stars identified as candidate or confirmed binaries.

Figure~\ref{fig:mass_EqW} reveals that H$\alpha$ emission is present in stars with masses up to $\approx$0.8~\Msun\ in M37, whereas in HyPra stars, this limit is $\approx$0.6 \Msun. The transition from a population of active and inactive stars at a given mass to one where all the stars are active also appears to occur at a higher mass in M37 than in HyPra ($\approx$0.5 v. $\approx$0.3 \Msun). 

Under the paradigm of chromospheric activity decaying with age, our EqW measurements confirm that M37 is younger than HyPra, as its $\approx$0.6$-$0.8 \Msun\ stars have not yet spun down enough to shut off their chromospheric \Ha\ emission.
 
Strong H$\alpha$ emission can be a sign of flaring. Unfortunately, our Hectospec spectra do not extend sufficiently into the blue to see the rise in the continuum and higher-order Balmer emission lines that are typical of flare spectra \citep[e.g., figure 8 of][]{Agueros2009}. However, our inspection of the continuum between 5000 and 6000~\AA\ suggests it is unlikely that we caught any of these stars during a significant flare.

For all M37 stars with \Ha\ emission, we derive \LLH.  This quantity is frequently used to compare levels of chromospheric activity between stars in samples spanning a range of masses, as it indicates the significance of \Ha\ flux relative to the entire energy output of the star. \LLH\ can be obtained by using the relation
\begin{equation}
    L_{\mathrm{H}\alpha}/L_{\mathrm{bol}} = -W_{\mathrm{H}\alpha} \frac{f_0}{f_{\mathrm{bol}}},
\end{equation}
where $W_{\mathrm{H}_{\alpha}}$ is the \Ha\ EqW, $f_0$ is the continuum flux near the \Ha\ line, and $f_{\mathrm{bol}}$ is the apparent bolometric flux.

For non-flux-calibrated spectra such as our Hectospec spectra, an alternative approach is to calculate $\chi = f_0 / f_{\mathrm{bol}}$ as a function of color. D14 derived an empirical relation between $\chi$ and color using spectra from the \textsc{Phoenix Aces} model spectra \citep{husser2013} with solar metallicity, log($g$) = 5.0, and 2500 $\leq T_{\mathrm{eff}} \leq$ 5200. We use that relation to derive $\chi$ values for our M37 stars by linearly interpolating between the D14 colors. For stars with a 2MASS counterpart, we use their de-reddened $(i-J)$ colors (requiring that the photometric quality be ``C'' or better); for stars with no 2MASS counterparts, we use de-reddened $(g-i)$ instead.\footnote{We derive $giJ$ extinction values using the tables of \citet{schlafly2011}, assuming $R_V$ = 3.1 and adopting a reddening $E$($B-V$) = 0.227 (H08).} 

\subsection{Accounting for Contamination by an Emission Nebula}\label{contamination}
Some of the Hectospec spectra exhibit [N\textsc{ii}]$\lambda$6548 and $\lambda$6584 \AA\ emission. [N\textsc{ii}] emission is known to originate in warm gas clouds, with temperatures near 10$^4$ K and atomic hydrogen densities around 10$^2-10^4$ cm$^{-3}$.

It is unlikely that the M37 stars are responsible for this emission, as they are not luminous enough to generate Str\"omgren spheres \citep{Stromgren39}. A more likely source is a foreground emission nebula, given the apparent extent of the emission near the core of M37 and the strength of the [N\textsc{ii}] lines with respect to that of the stellar spectra. Unfortunately, our Hectospec spectra do not have enough wavelength resolution to perform velocity measurements, and we are not able to measure radial velocity differences between the nebula and the M37 stars.

\begin{figure*}
\centering
\includegraphics[scale=0.65]{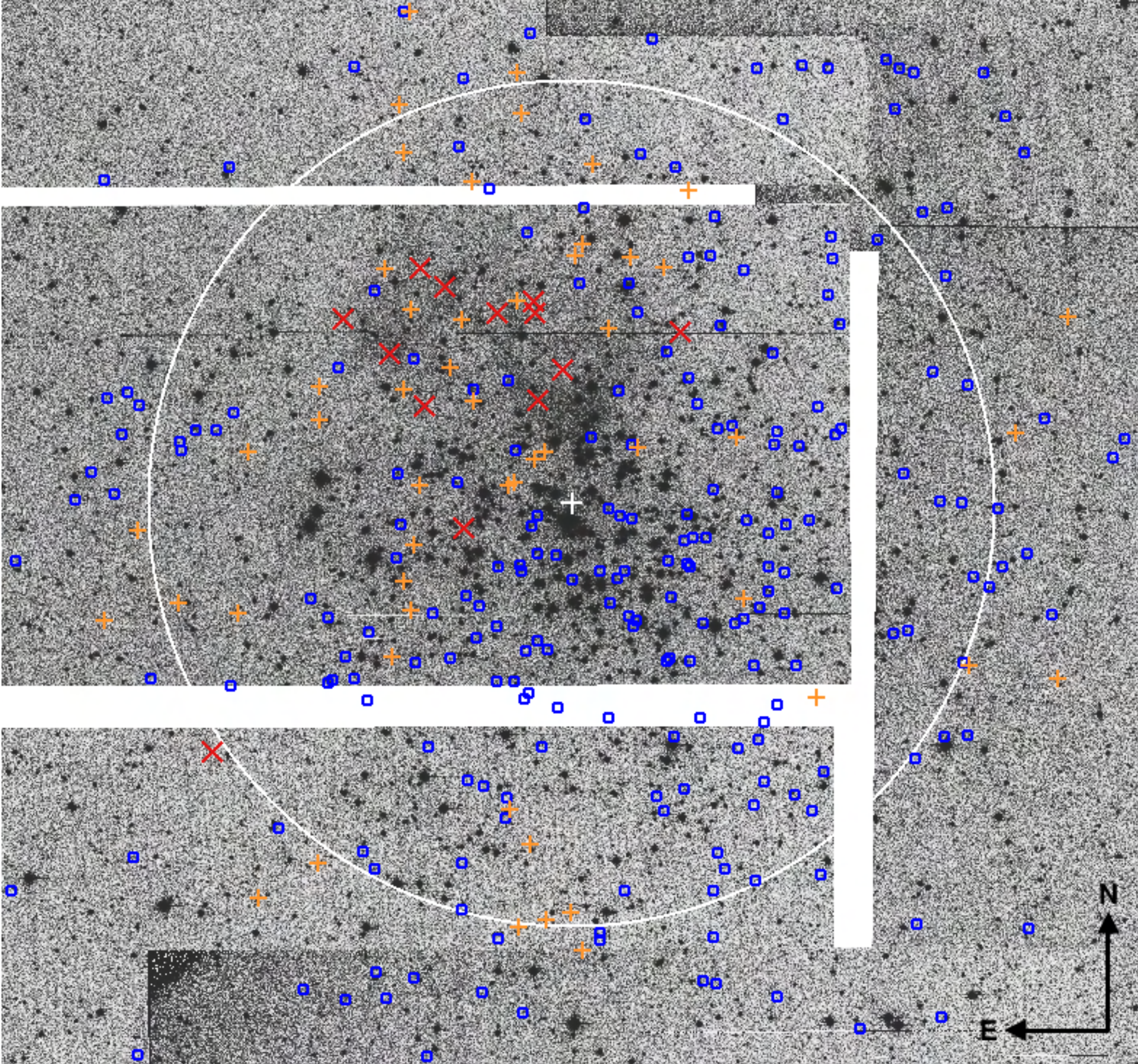}
\caption{Mosaic of IPHAS \Ha\ images of M37 spanning 0.52 degrees in RA and 0.42 in DEC. North is up and East is to the left. To provide a sense of scale, the white circle marks the 10$\arcmin$ radius from the cluster center (white cross). Red x symbols indicate spectra with [N\textsc{ii}]$\lambda$6584 EqW $<-3.0$, orange crosses, spectra with $-3.0\ \leq$ [N\textsc{ii}]$\lambda$6584 EqW $<-0.5$, and blue circles, spectra with $-0.5\leq$ [N\textsc{ii}]$\lambda$6584 EqW $\leq0$.}
\label{fig:iphas}
\end{figure*}

\begin{figure}
\includegraphics{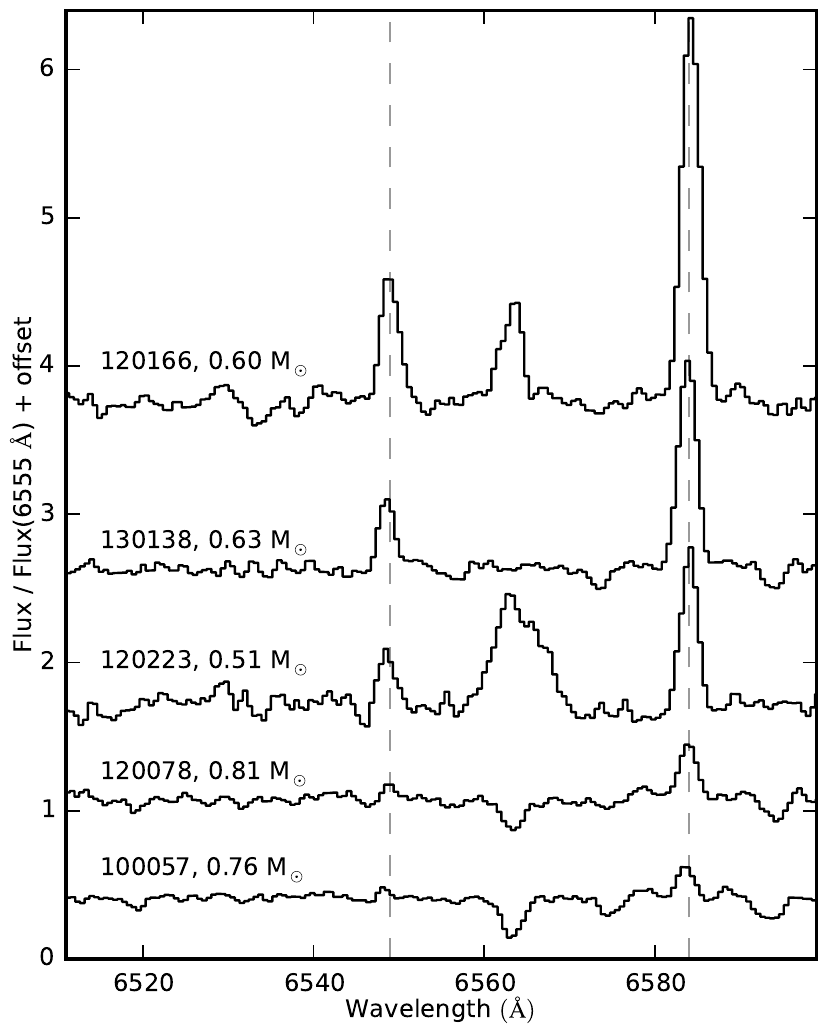}
\caption{Five M37 stars where [N\textsc{ii}] emission is observed, labeled with their H08 designations and photometrically derived masses. Each spectrum is normalized to the flux at 6555 $\AA$ and smoothed using a four pixel smoothing window. The dashed vertical lines indicate the [NII]$\lambda$6548 and $\lambda$6584 \AA\ doublet.}
\label{fig:NII}
\end{figure}

We measured the EqW of the [N\textsc{ii}] emission lines using the same technique described in Section~\ref{Halpha}. Figure~\ref{fig:iphas} shows a mosaic of \Ha\ images from the Isaac Newton Telescope Photometric \Ha\ Survey (IPHAS) of the Northern Galactic Plane \citep{Drew2005} covering the core of M37. A white cross indicates the center of the cluster. For a sense of scale, we draw a white circle marking a 10$\arcmin$ radius from the cluster center. Spectra with [N\textsc{ii}]$\lambda$6584 EqW $<-3.0$ are indicated with red x symbols; spectra with $-3.0 \leq $ [N\textsc{ii}]$\lambda$6584 EqW $\leq 0.5$, with orange crosses; and spectra with $-0.5 < $ [N\textsc{ii}]$\lambda$6584 EqW $\leq0$, with blue circles. Most spectra with strong [N\textsc{ii}] emission coincide with the location of a dim nebular structure (northeast of the cluster center). Five examples of our spectra with [N\textsc{ii}] emission lines are shown in Figure~\ref{fig:NII}. 

Since the ionization potential energy of N is very similar to that of H, \Ha\ emission in principle always accompanies [N\textsc{ii}] emission in emission nebulae. This means that some of our \Ha\ EqW measurements for M37 stars may be contaminated by \Ha\ emission from the nebula.

Observations of many different planetary nebulae show that the intensity ratio [N\textsc{ii}]$\lambda$6584/\Ha\ can vary significantly, from 0 to $\approx$4 \citep[see, e.g.,][]{White1952, Kaler1983}, depending on the metal content and density of the nebula \citep{Draine2011}. It is, therefore, difficult to use this ratio alone to characterize the nature of the nebula \citep{Frew2010}. Furthermore, the [O\textsc{iii}]$\lambda$4959 and $\lambda$5007 \AA\ lines that are also typical of emission nebulae, and that are useful in determining gas density and temperature, fall outside the wavelength range of our Hectospec spectra, thus preventing us from characterizing the nebula. We cannot determine  the amount of nebular H$\alpha$ emission contaminating our stellar spectra. We therefore exclude cluster stars with strong [N\textsc{ii}] emission ([N\textsc{ii}]$\lambda6584$ EqW $<-3.0$, henceforth referred simply as [N\textsc{ii}] contamination) from our analysis.

\subsection{Calculating Rossby Numbers}\label{rossby}
\citet{Noyes1984} found that to study the relation between rotation and activity, using $R_o = P_{\mathrm{rot}} / \tau$, where $\tau$ is the convective turnover time, rather than \Prot\ directly yields easier-to-interpret results, as it removes the mass dependence observed in stars with saturated levels of activity. In Paper I we estimated $\tau$ values for M37 members using the empirical stellar mass-$\tau$ relation of \citet{wright2011}, which is based on \Prot\ and \LX\ measurements of over 800 stars in the mass range 0.09$-$1.36~\Msun.\footnote{We chose to use the fitted values of \citet{wright2011} based on $\beta=-2.0$ instead of $-2.7$, as we questioned in Paper I the validity of the ``unbiased'' sample from which these authors obtained $\beta=-2.7$.} With those $\tau$ values we calculated $R_o$ for all M37 members with \Prot. Here we use these same $\tau$ and $R_o$ values to study the rotation-activity relation in \Ha\ space.

\section{Results and Discussion}\label{results}

\subsection{Relationship Between \Ha\ Emission and Rotation}\label{Halpharotation}
To study the relationship between chromospheric activity and stellar rotation, we perform a similar analysis as in Paper I, where we parametrized the \LLX--$R_o$ relation as a flat region connected to a power-law. Here, we fit this same functional form to \LLH:
\begin{equation}\label{eq:rossby}
  \frac{L_{\mathrm{H\alpha}}}{L_{\mathrm{bol}}} = \left\{
  \begin{array}{l l}
    \left(\frac{L_{\mathrm{H\alpha}}}{L_{\mathrm{bol}}}\right)_{\mathrm{sat}} & \quad \textrm{if $R_o\le R_{o\mathrm{,sat}}$}\\
    C R_o^\beta & \quad \textrm{if $R_o$ > $R_{o\mathrm{,sat}}$}
  \end{array} \right.
\end{equation}
where (\LLH)$_{\mathrm{sat}}$ is the activity saturation level, $R_{o\mathrm{,sat}}$ is the saturation threshold, $\beta$ is the power-law index characterizing the unsaturated regime, and $C$ is a constant. We fit this model to the 65 single cluster members that have both $R_o$ and \LLH\ measurements and do not have significant [N\textsc{ii}] contamination.

We convert the model in Equation \ref{eq:rossby} into log-space for the fit and assume flat priors over all three parameters. We use the Markov-chain Monte Carlo (MCMC) package \texttt{emcee}\footnote{\url{http://dan.iel.fm/emcee/current/}} \citep{Foreman13} to carry out the fit. \texttt{emcee} builds posterior probability distributions for each parameter by performing a random walk in parameter space. This same approach was used in Paper I and by D14 for their sample of HyPra \Ha\ emitters. We use 300 initial seeds and 3000 iterations, one tenth of which were to burn-in the walkers.

\begin{figure}
\includegraphics{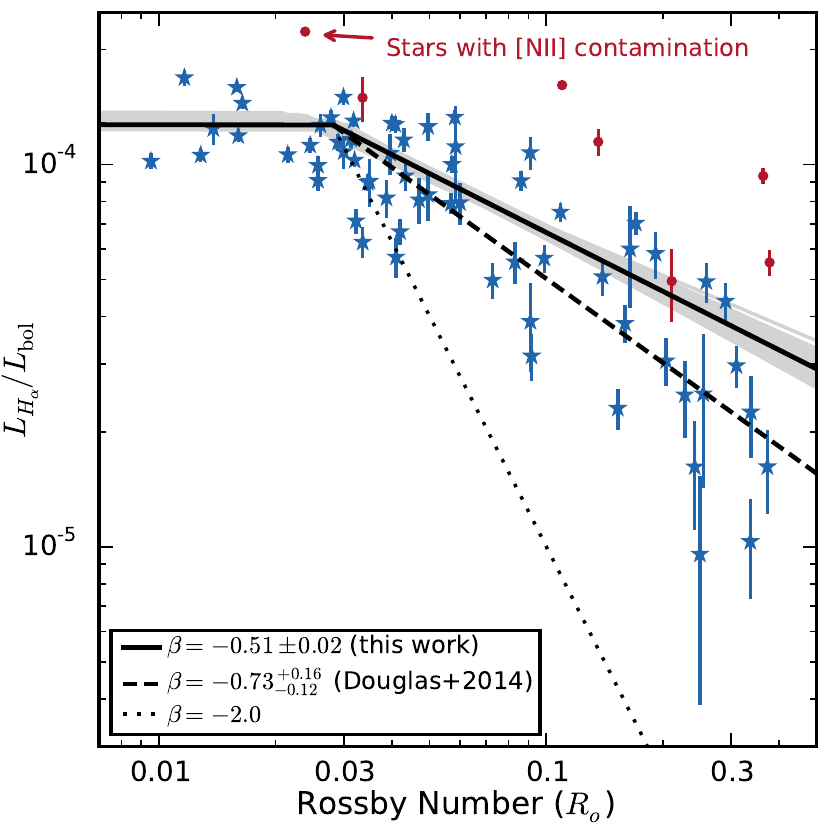}
\caption{\LLH\ as a function of Rossby number $R_o$ for M37 single cluster members. The solid black line is the maximum $a\ posteriori$ fit from the MCMC algorithm for the model described by Equation~\ref{eq:rossby}. The gray lines are 200 random samples from the posterior probability distributions. The dashed line indicates the fit found by D14 on a sample of HyPra stars. The dotted line indicates a slope of $-2.0$, which is typically found in studies of X-ray activity indicators.
Red circles indicate cluster stars with [N\textsc{ii}] contamination ([N\textsc{ii}]$\lambda6584$ EqW $<-3.0$, see Section~\ref{contamination}); these stars are excluded from our analysis.}
\label{fig:Rossby}
\end{figure}

\begin{figure}
\includegraphics{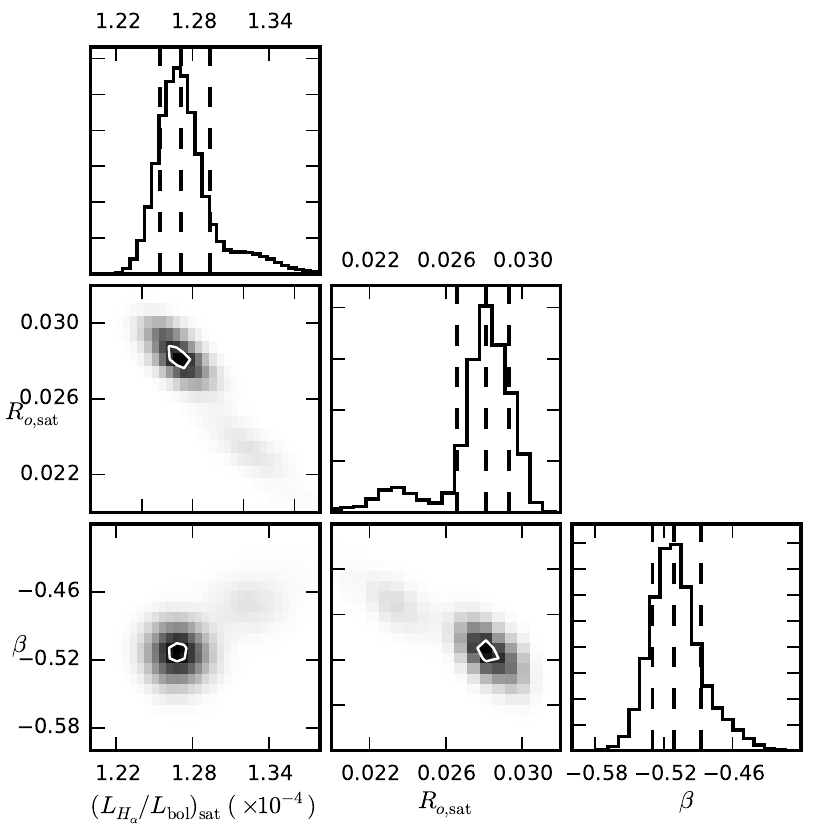}
\caption{Marginalized posterior probability distributions from the MCMC analysis using \texttt{emcee}. The parameter values of the \textit{a\ posteriori} model are the peaks of the one-dimensional distributions; the vertical dashed lines approximate the median and 68-percentiles. The two-dimensional distributions illustrate covariances between parameters; the white contour lines approximate the 68-percentile of the distributions.}
\label{fig:correlations}
\end{figure}

Shown in gray lines in Figure~\ref{fig:Rossby} are 200 models drawn at random from the posterior probability distributions. The solid black line indicates the most probable model, i.e., the maximum $a\ posteriori$ model. The resulting best-fit parameters are (\LLH)$_{\mathrm{sat}}$ = (1.27$\pm0.02$)$\times 10^{-4}$, $R_{o\mathrm{,sat}}=0.03\pm$0.01, and $\beta=-0.51\pm$0.02. The parameter values correspond to the 50$^{\rm th}$ quantile, and the uncertainties correspond to the 16$^{\rm th}$ and the 84$^{\rm th}$ quantiles (for consistency with $1\sigma$ Gaussian uncertainties). We highlight in Figure~\ref{fig:Rossby} single cluster members with strong [N\textsc{ii}] contamination, but we exclude them from the fit analysis. Figure~\ref{fig:correlations} shows the posterior probability distributions for all parameters.

\subsubsection{Sensitivity of Results to Choices of \Pmem\ and {\rm [N\textsc{ii}]} Contamination Thresholds}
We test how sensitive our MCMC results are to the two parameters used to define our sample of stars, namely, \Pmem\ and [N\textsc{ii}] contamination. We re-run our fit using a more conservative \Pmem\ = 0.7 cutoff instead of our adopted 0.2, and also adopting a more conservative [N\textsc{ii}]$\lambda6584$ EqW $=-1.0$ cutoff instead of $-3.0$. In both cases, the new $R_{o\mathrm{,sat}}$ and $\beta$ values are within 1$\sigma$ of our original best-fit results, underlining their reliability.

\subsubsection{Comparison to Previous Results}
The saturation level (\LLH)$_{\mathrm{sat}} = (1.27$$\pm$$0.02) \times 10^{-4}$ for M37 stars is the same, within the errors, as the one found by D14 for HyPra stars, (1.26$\pm$$0.04)\times 10^{-4}$. The saturated regime includes stars with $R_o$ numbers up to 0.03$\pm0.01$. This saturation threshold is smaller by at least a factor of two than the typical value found in other sets of stars. For example, D14 found $R_{o\mathrm{,sat}}=0.11^{+0.02}_{-0.03}$ using H$\alpha$ activity measurements of HyPra stars, \citet{Jackson2010} found $R_{o\mathrm{,sat}}\approx0.1$ using Ca\textsc{ii} activity measurements of NGC~2516 stars, and \citet{Randich2000b} found $R_{o\mathrm{,sat}}\approx0.16$ using X-ray activity measurements of a heterogeneous sample of field and cluster stars. More surprisingly, in Paper I we found that $R_{o\mathrm{,sat}}=0.09\pm0.01$ for our sample of X-ray-emitting M37 stars.

In the unsaturated regime, \Ha\ activity decreases as a power-law with slope $\beta=-0.51$$\pm$0.02. This slope is statistically shallower than the $\beta=-0.73^{+0.16}_{-0.12}$ that D14 found for HyPra stars. In turn, the $\beta$ values found in these two studies of \Ha\ activity are significantly shallower than the values found in studies of X-ray activity. For example, \citet{guedel1997} found $\beta=-2.64$$\pm$0.12 for a sample of 12 solar-type stars of ages 0.07 to 9 Gyr, while \citet{Randich2000b} and \citet{wright2011} found $\beta=-2.10$$\pm$0.09 and $-2.18$$\pm$0.16, respectively, for X-ray samples of field and cluster stars. And in Paper I we found $\beta=-2.03_{-0.14}^{+0.17}$ in our {\it Chandra} study of M37.

\subsection{Chromospheric versus Coronal Activity}\label{chromosphericcoronal}
A difference between the decay of \Ha\ and X-ray emission has been observed before \citep[e.g.,][]{Hodgkin1995, Preibisch2005a, stelzer2013}. Our results confirm this discrepancy and highlight a physical distinction between the decay rate of chromospheric activity versus coronal activity in low-mass stars. Furthermore, since $R_{o\mathrm{,sat}}$ is smaller for \LLH\ versus $R_o$ than for \LLX\ versus $R_o$, our results indicate that as stars spin down, their chromospheres exit the saturated regime before their coronae do. This leads us to examine our \Ha\ and X-ray data in more detail. 

\begin{figure}
\includegraphics{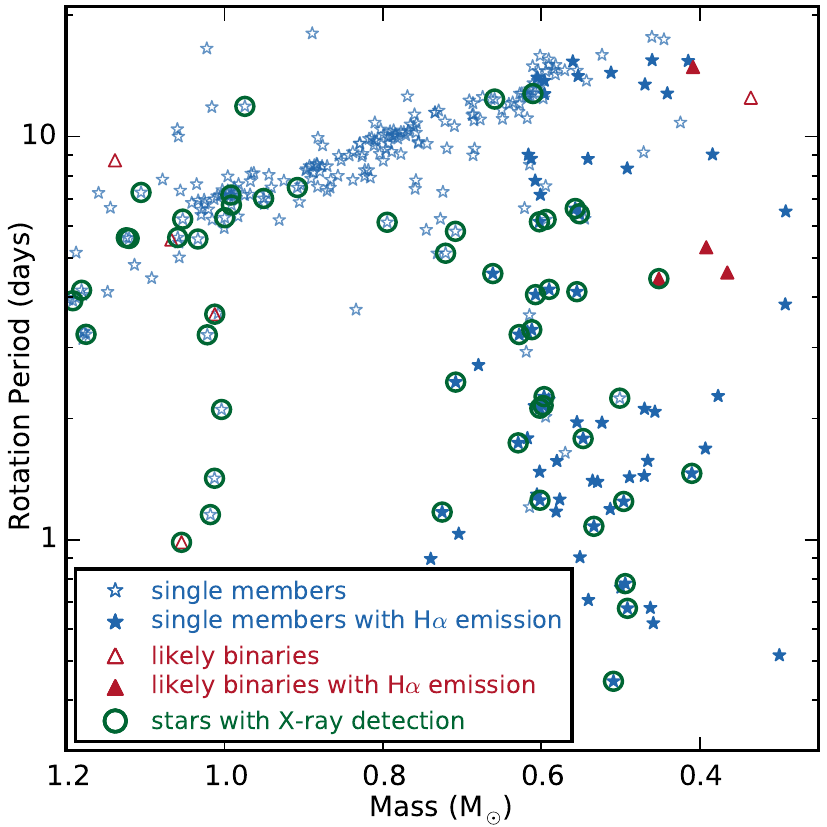}
\caption{\Prot\ versus mass for M37 members. Single stars  are indicated with blue symbols, and likely binaries with red triangles. Stars with \Ha\ emission have their symbols filled. Stars with X-ray emission identified in Paper I are highlighted with a green circle.}
\label{fig:massperiods}
\end{figure}

Figure~\ref{fig:massperiods} shows \Prot\ versus stellar mass for M37 stars, including both single (blue star symbols) and likely binary members (red triangles). Stars with detected \Ha\ emission have their symbols filled, and stars with detected X-rays are highlighted with a green circle. In total there are 50 stars with both \Ha\ and X-ray emission (nine are likely binaries), all with masses $\leq$0.73 \Msun; of these, 24 have \Prot\ measurements.

Most active low-mass stars should have some level of \Ha\ emission. The fact that we do not detect this emission for stars with masses $>$0.73 \Msun\ in our M37 sample may be because \Ha\ absorption, which increases in intensity with stellar mass \citep{Cram1985}, dominates the H$\alpha$ spectral feature in those stars. In principle it is possible to separate the levels of \Ha\ emission and absorption, but our spectra do not have the necessary resolution to perform this exercise. We therefore concentrate on stars for which we measure \Ha\ emission using the approach described in Section~\ref{Halpha}.

\begin{figure*}
\includegraphics{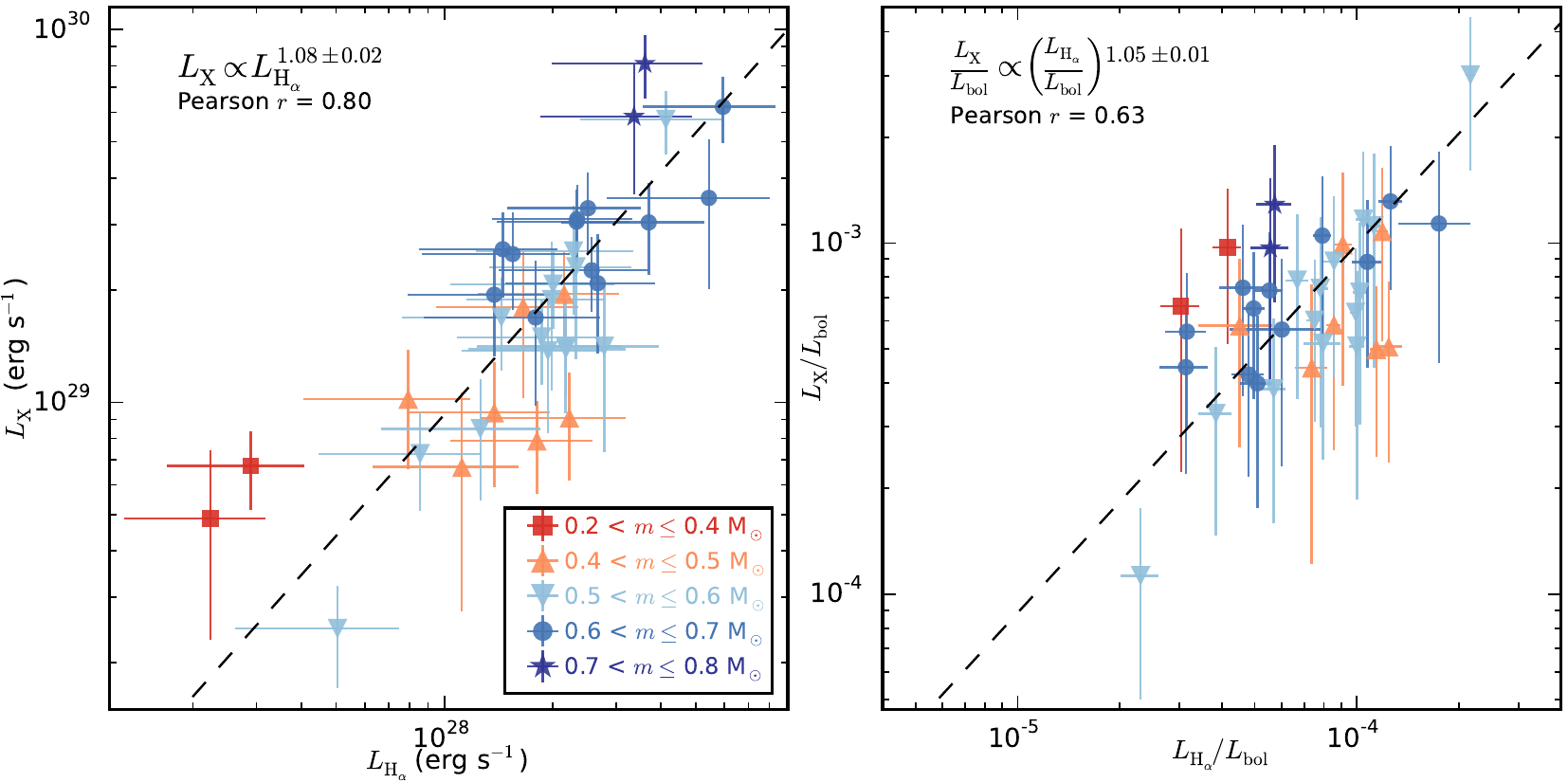}
\caption{{\it Left}: $L_X$ versus \LHa\ for single members of M37. {\it Right}: \LHa\ as a fraction of \Lbol\ versus $L_X$ as a fraction of \Lbol\ for the same stars. In both panels, red stars have masses between 0.2 and 0.4 \Msun, orange stars between 0.4 and 0.5, light blue stars between 0.5 and 0.6, blue stars between 0.6 and 0.7, and dark blue stars between 0.7 and 0.8. In both panels, the power-law relation and the Pearson $r$ found with a least squares bisector regression is annotated on top and indicated with a black dashed line.}
\vspace{0.3cm}
\label{fig:Ha_v_X}
\end{figure*}

Figure~\ref{fig:Ha_v_X} compares the activity of M37 stars ---excluding likely binaries and stars with [N\textsc{ii}] contamination--- in \Ha\ and in X rays. We use the 0.1--2.4 keV \LX\ values derived in Paper I. In the left panel, where \LHa\ is plotted versus \LX, there is a clear positive correlation between the two indicators of activity in our co-eval sample of M37 stars, in spite of the non-contemporaneous nature of the observations. A least squares bisector regression\footnote{The least squares bisector regression treats both the dependent and independent variables symmetrically. Therefore, it is the most appropriate method to use in this case, since we are interested in the underlying theoretical relationship between the two quantities \citep[see][]{Isobe1990}.} indicates that \LX\ is proportional to \LHa\ following a power-law such that \LX\ $\propto$ \LHa$^\alpha$, where $\alpha=1.08$$\pm$0.02 (black dashed line). The Pearson correlation coefficient $r=0.80$ for this relation suggests that the two quantities are strongly correlated.

\citet{Martinez2011} found a steeper power-law slope of 1.60$\pm0.07$ for a sample of F- to M-type stars. Using only M-dwarfs from that same study, \citet{stelzer2012} found a power-law slope of 1.29$\pm$0.15. Similarly, \citet{stelzer2013} found a steeper power-law slope of 1.61$\pm$0.23 for a sample of nearby (<10 pc) M dwarfs. \citet{pace2004}, on the other hand, did not find a correlation between \LX\ and chromospheric activity, the latter measured as the flux of H and K lines of Ca\textsc{ii} (their figure 8, left panel), for a sample of HyPra stars. 

In Figure~\ref{fig:Ha_v_X} we color-code stars by mass. The left panel shows a clear dependence of \LHa\ and \LX\ on mass. This is not surprising, since surface area increases with mass along the main sequence. In the right panel we divide both luminosities by \Lbol\ to remove this mass dependence. Once we do this, the \Ha--X-ray activity correlation is not as clear: a least squares bisector regression gives $\alpha = 1.05$$\pm$0.01. Its Pearson $r=0.63$ indicates that the \LLH--\LLX\ correlation is not as strong as with \LHa--\LX. Furthermore, if we exclude the two outliers in this panel (the two 0.5--0.6 \Msun\ stars well above/below the main locus), $\alpha$ decreases to 0.89$\pm$0.01 and the Pearson $r$ decreases to 0.43.

\begin{figure}
\includegraphics{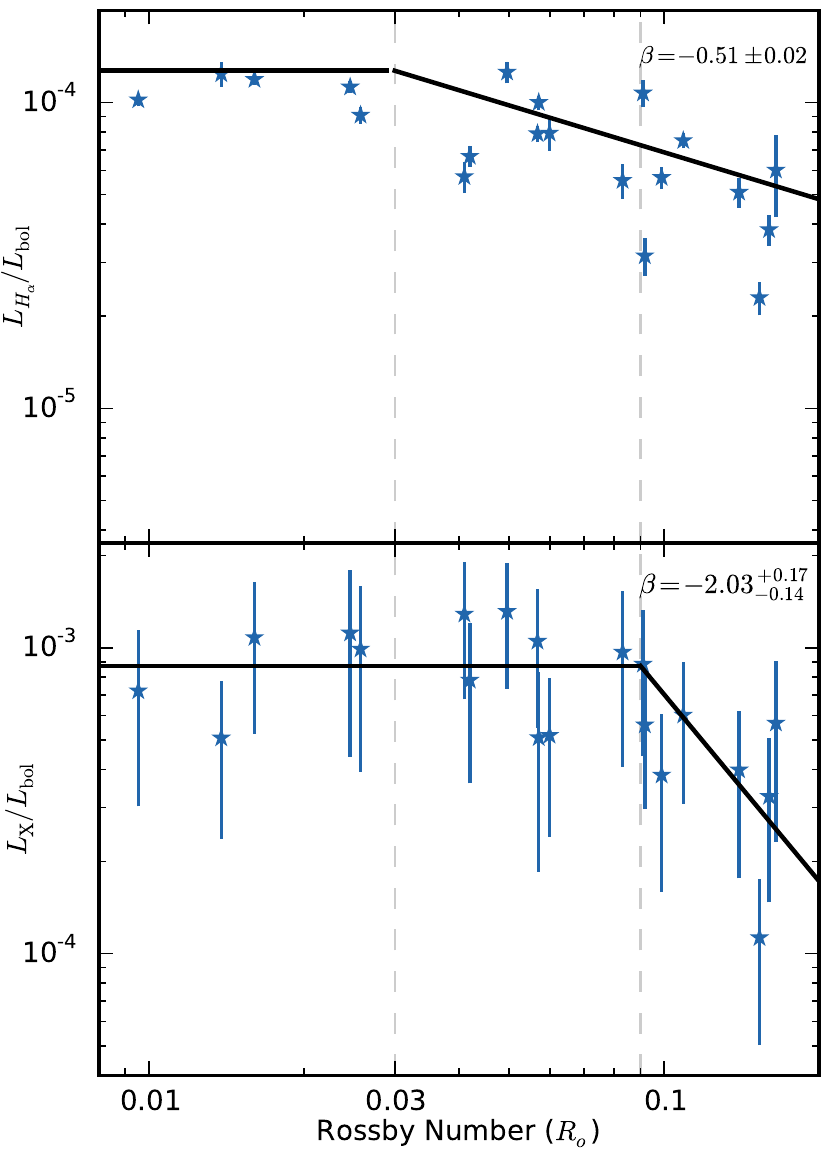}
\caption{\LLH\ v. $R_o$ (top) and \LLX\ v. $R_o$ (bottom) for a sample of M37 stars with both \LHa\ and \LX\ measurements. Solid lines are the MCMC best-fit parameters (top, for Equation~\ref{eq:rossby}; bottom, for equation 2 of Paper I). Vertical lines mark the two threshold ${R_{o\mathrm{,sat}}}$ values. The power slopes $\beta$ are noted in each panel.}
\label{fig:Rossby_HaX}
\end{figure}

This apparent stagnation of \LLX\ with respect to changes in \LLH\ is more evident in Figure~\ref{fig:Rossby_HaX}, in which both panels include only M37 stars with both \LHa\ and \LX\ measurements. The vertical dashed lines indicate the two saturation thresholds: ${R_{o\mathrm{,sat}}} = 0.03$ for the \LLH--$R_o$ relation, and ${R_{o\mathrm{,sat}}} = 0.09$ for \LLX--$R_o$ (from Paper I). Coronal emission saturates at slower \Prot\ than chromospheric emission.

A larger ${R_{o\mathrm{,sat}}}$ value for \LLX\ can be interpreted as evidence for coronal stripping. This mechanism is sometimes invoked to explain the saturation level observed in \LLX, in which centrifugal forces strip the outermost layers of the corona \citep[e.g.,][]{Christian2011, Argiroffi2016}. It is possible that coronal stripping is also partially driven by an imbalance between magnetic and plasma pressure equilibrium, in a scenario with highly energetic stellar winds \citep[e.g.,][]{Parker1960, Jardine1999}.

The potential coronal stripping seen in M37 stars may point to the fact that coronal indicators such as X rays present limitations as tracers of stellar activity. Tracers that originate closer to the stellar surface (e.g., measurements of photospheric or chromospheric emission) are perhaps more representative of the magnetic heating mechanism in the stellar atmosphere. Indeed, \citet{davenport2016} found a similar $R_{o\mathrm{,sat}} \approx 0.03$ in a study of stellar flares measured from {\it Kepler} light curves of $\approx$290 stars. Furthermore, studies including other chromospheric tracers (e.g., Ca\textsc{ii} or Mg\textsc{ii} emission) find no evidence for supersaturation \citep[e.g.,][]{Christian2011}, supporting the idea that stripping mechanisms do not affect the inner atmospheric layers.

It would be valuable to study other indicators of chromospheric activity for the same sample of M37 stars. The correlation between \Ha\ emission and other chromospheric indicators has been found to be ambiguous at best \citep[e.g.,][]{Strassmeier1990, Scandariato2016}. However, if $R_{o\mathrm{,sat}}$ for other chromospheric indicators agreed with the one we find for \Ha, it would highlight the limitation of X rays as a reliable estimator of stellar activity. Chromospheric indicators may have more predictive power in the parametrization of the rotation-activity relation and, ultimately, the ARAR.

\section{Summary}\label{summary}
We present the results of a spectroscopic survey to characterize chromospheric activity, as measured by \Ha\ emission, in low-mass members of the 500-Myr-old open cluster M37. We measured H$\alpha$ EqWs for 294 cluster members, 125 of which show the line in emission. 

We use properties previously cataloged for M37's members in Paper I, including $P_{mem}$, mass, \Lbol, and $\tau$, and $P_{rot}$ measurements from \citet{messina08a} and \citet{Hartman09} to examine the dependence of activity on rotation in this cluster.

An emission nebula appears to have contaminated a small fraction of our  spectra, namely in the form of strong [N\textsc{ii}]$\lambda$6548 and $\lambda$6584 \AA\ emission lines, and potentially some \Ha\ emission as well. We exclude from our rotation-activity analysis cluster stars for which this contamination is strong.

M37 stars exhibit \Ha\ emission in stars as massive as $\approx$0.8~\Msun, compared to $\approx$0.6~\Msun\ in the $\approx$650-Myr-old merged Hyades/Praesepe sample of stars \citep{Douglas14}. This confirms that M37 is younger than these two clusters.

We identify saturated and unsaturated regimes in the dependence of \LLH\ on $R_o$. Rotators with $R_o$ less than the saturation threshold $R_{o,\mathrm{sat}}= 0.03\pm$0.01 are saturated, and converge to an activity level of \LLH\ = $(1.27\pm0.02)\times10^{-4}$. This $R_{o,\mathrm{sat}}$  is statistically smaller than the canonical $R_o=0.1$ found in most studies of the rotation-activity relation. Only the study of flaring M dwarfs by \citet{davenport2016} finds a comparably small $R_{o,\mathrm{sat}}$.

In the unsaturated regime, faster rotators have increasing levels of chromospheric activity, with \LLH($R_o$) following a power-law of index $\beta=-0.51\pm$0.02, slightly shallower than the one found by \citet{Douglas14} for Hyades/Praesepe stars. We confirm previous findings that show chromospheric activity decaying at a much slower rate than coronal activity with increasing $R_o$. 

While a comparison of \LHa\ and \LX\ for M37 stars with measurements of both reveals a close to 1:1 relation, removing the mass-dependencies by comparing instead \LLH\ and \LLX\ does not provide clear evidence for a linear relation.
This indicates that chromospheric and coronal activity indicators may be interchangeable in activity studies, but only in their pure luminosity forms.

We find that $R_{o,\mathrm{sat}}$ is smaller for our chromospheric indicator than for our coronal indicator of activity ($R_{o,\mathrm{sat}}=0.03\pm$0.01 versus 0.09$\pm$0.01). We interpret this as possible evidence for coronal stripping, likely a result of both centrifugal forces and an imbalance between magnetic and plasma pressure equilibrium.

\acknowledgments We thank the SAO Pre-doctoral Program for hosting A.N.~for three months at the Harvard Smithsonian Center for Astrophysics in Cambridge, MA. We thank the anonymous referee for comments that improved the paper. M.A.A.\ acknowledges support provided by the NSF through grant AST-1255419. Support for this work was provided by the National Aeronautics and Space Administration through Chandra Award Number GO6-17039X issued by the Chandra X-ray Observatory Center, which is operated by the Smithsonian Astrophysical Observatory for and on behalf of the National Aeronautics Space Administration under contract NAS8-03060.

This paper makes use of data obtained as part of the INT Photometric \Ha\ Survey of the Northern Galactic Plane (IPHAS, \url{www.iphas.org}) carried out at the Isaac Newton Telescope (INT). The INT is operated on the island of La Palma by the Isaac Newton Group in the Spanish Observatorio del Roque de los Muchachos of the Instituto de Astrofisica de Canarias. All IPHAS data are processed by the Cambridge Astronomical Survey Unit, at the Institute of Astronomy in Cambridge. The bandmerged DR2 catalogue was assembled at the Centre for Astrophysics Research, University of Hertfordshire, supported by STFC grant ST/J001333/1.

\setlength{\baselineskip}{0.6\baselineskip}
\bibliography{references}
\setlength{\baselineskip}{1.667\baselineskip}

\end{document}